\documentclass[%
 aip,
 amsmath,amssymb,
 reprint,%
]{revtex4-1}

\usepackage{graphicx}
\usepackage{dcolumn}
\usepackage{bm}

\usepackage[utf8]{inputenc}
\usepackage[T1]{fontenc}
\usepackage{mathptmx}
\usepackage{etoolbox}
\usepackage[separate-uncertainty=true]{siunitx}

\usepackage{physics}
\usepackage{upgreek} 
\usepackage{booktabs}
\usepackage{float}

\makeatletter
\def\@email#1#2{%
 \endgroup
 \patchcmd{\titleblock@produce}
  {\frontmatter@RRAPformat}
  {\frontmatter@RRAPformat{\produce@RRAP{*#1\href{mailto:#2}{#2}}}\frontmatter@RRAPformat}
  {}{}
}%
\makeatother
\begin{document}

\preprint{AIP/123-QED}

\title{Operating a bistable qubit}
\def\RLEaffil{Research Laboratory of Electronics, Massachusetts Institute of Technology, Cambridge, MA 02139, USA}
\def\UREGvaffil{Institute of Experimental and Applied Physics and Halle-Berlin-Regensburg Cluster of Excellence CCE, University of Regensburg, 93040 Regensburg, Germany}
\def\NTNUaffil{Department of Physics, Norwegian University of Science and Technology, NO-7491 Trondheim, Norway}
\def\QMaffil{Q.M Technologies Ltd. (Quantum Machines), Tel Aviv-Yafo, Israel}
\def\Qolaffil{Qolab, Madison, Wisconsin 53706, USA}

\author{Fabrizio~Berritta}
\email{fabrizio.berritta@mit.edu}
\affiliation{\RLEaffil}
\affiliation{\UREGvaffil}
\author{Jan~A.~Krzywda}
\affiliation{Lorentz Institute for Theoretical Physics \& Leiden Institute of Advanced Computer Science, Universiteit Leiden, 2311 EZ Leiden, The Netherlands}
\author{Tom~Dvir}
\affiliation{\QMaffil}
\author{Paul~Buttles}
\affiliation{\Qolaffil}
\author{Stanislav~Eilhart}
\affiliation{\Qolaffil}
\author{Jeroen~Danon}
\affiliation{\NTNUaffil}
\author{Ferdinand~Kuemmeth}	
\email{ferdinand.kuemmeth@ur.de}
\affiliation{\UREGvaffil}
\affiliation{QM Technologies GmbH, 60596 Frankfurt am Main, Germany}
\affiliation{Niels Bohr Institute, University of Copenhagen, 2100 Copenhagen, Denmark}

\def\fL{f_{\text{L}}}
\def\fH{f_{\text{H}}}
\def\fc{f_\text{c}}
\def\fq{f_\text{q}}

\date{\today}

\begin{abstract}
Parasitic two-level-system (TLS) defects limit the stability and performance of solid-state quantum processors. Their interaction with a qubit can cause discrete, stochastic shifts of the qubit frequency, making the qubit bistable. We experimentally demonstrate an adaptive protocol for operating a bistable qubit with high fidelity using a classical controller powered by a field-programmable gate array (FPGA). Our ``1-bit feedback'' protocol estimates the qubit's bistable frequency and corrects the control frame using only one single-shot measurement, reaching the information limit set by the qubit's intrinsic entropy. We validate the protocol in a superconducting qubit by suppressing TLS-induced Ramsey beating, and deploy it to stabilize gate fidelities over time with approximately $\SI{136}{\kilo\hertz}$ estimation bandwidth and a $77\%$ error reduction. Our approach provides a simple, yet fundamentally efficient strategy for mitigating dephasing errors induced by strongly coupled TLS defects, and may enable the operation of large future qubit arrays suffering from few remaining, discrete instabilities. 
\end{abstract}

\maketitle

\section{Introduction}

Two-level system (TLS) defects are one of the major contributors to decoherence in superconducting qubits and semiconductor spin qubits, though their detailed microscopic origins remain unknown.\cite{muller2019, Siddiqi2021,Murray2021,rojas2023spatial,ye2024, donnelly2025noise, Rojas2026} TLSs can induce energy relaxation from the excited to the ground state of the qubit, and they can also generate slowly varying noise that contributes to qubit dephasing. An ensemble of TLSs gives rise to a $1/f$-like charge or flux noise power spectrum at sub-Hz to kHz timescales, whereas a single strongly coupled TLS can cause discrete, telegraphic shifts of the qubit frequency. TLSs can thus be non-Markovian noise sources, which exhibit memory effects and introduce significant overhead for error mitigation.\cite{hakoshima2021relationship, kam2024detrimental,gao2026non} Their dynamical decoupling is not universally effective and may not align with specific experimental goals.\cite{gustavsson2012dynamical, szankowski2017environmental} In order to realize fault-tolerant quantum computing with error-corrected solid-state qubits, non-Markovian noise likely needs to be reduced.\cite{pataki2024coherent} 

The detrimental effects of TLSs can be mitigated by improving the device design, material and fabrication processes, or by controlling the noise source itself.\cite{Siddiqi2021,mcrae2021,weeden2025, wolff2026structural,grabovskij2012,lisenfeld2019,bilmes2020, kim2025error, chen2025, dane2025, ye2025stabilizing, degnan2026reducing, roy2026two} Here we focus on low-latency estimation of a telegraphically fluctuating qubit parameter: Several online (during experimental data collection) estimation methods have been proposed to enhance calibration efficiency and error mitigation against $1/f$-noise drifts in stochastic qubit parameters, but they do not directly address bistable qubits whose frequency switches between discrete values.\cite{Gebhart2023,Arshad2024, Berritta2024a, dumoulin2024silicon, Berritta2025_FBS, marciniak2026} This work demonstrates an information-limited real-time (on the qubit coherence timescale) estimation approach applied to a bistable qubit.
 
In superconducting qubits, TLSs across the GHz spectrum, located in the aluminum oxide or other dielectric material, can couple strongly to the qubit by electric-dipole interaction. They can also indirectly limit the fidelity of quantum operations by shortening qubit relaxation times.\cite{Malley2015} Switches in relaxation time by an order of magnitude caused by TLSs have been observed over timescales of just milliseconds.\cite{Berritta2025_T1}
TLSs can also reduce gate fidelities by inducing qubit-frequency fluctuations, and sometimes by randomly switching between two or three values;\cite{muller2019, Siddiqi2021,Murray2021} here we focus on the case of bistability between two known qubit frequencies.

Our efficient ``1-bit feedback'' protocol is executed on a bistable superconducting qubit by programming a commercial controller with an integrated field-programmable gate array (FPGA) to calibrate the qubit control frame in real time using only a single-shot measurement outcome, in the presence of slow telegraphic fluctuations that are likely due to a strongly coupled TLS.\cite{Krantz2019, Blais2021}  
Single-shot Ramsey discrimination of discrete qubit-frequency states has previously been demonstrated in transmon charge-parity detection, with feedback resetting the qubit state for repeated measurements.\cite{Rist2013} Here, we use each single-shot outcome to update the control frame of subsequent quantum operations in real time, suppressing TLS-induced beating in Ramsey fringes and by eliminating telegraphic fluctuations of quantum gate infidelities, resulting in a significant enhancement of the resulting time-averaged gate fidelity.
Our scheme demonstrates intermittent and efficient calibration of a bistable qubit, making it ideal for stable quantum circuit execution in the presence of telegraphic drifts in the qubit frequency.

We use a transmon qubit at the Israeli Quantum Computing Center (IQCC), controlled by a commercial low-latency FPGA-based controller (Quantum Machines OPX1000). Device and setup details are provided in Appendix~\ref{app:setup_device}. Our transmon with anharmonicity $\approx \SI{-300}{\mega\hertz}$ and $E_{\rm J}/E_{\rm c}\approx 53$ is observed to stochastically switch between two configurations (``modes'') that differ in the qubit transition frequency. This is likely caused by some effective coupling between the transmon and a slow TLS (slow relative to the time scale of initializing, operating, reading out the transmon and updating control pulse parameters on the controller).  This TLS-induced switching between two modes makes the qubit \textit{bistable}: whenever the TLS changes configuration, the qubit frequency switches between two known discrete values. In the high-frequency mode (H mode), the qubit frequency is $\approx \SI{5.10}{\giga\hertz}$, whereas in the low-frequency mode (L mode), the qubit frequency $\fq$ decreases by $\Delta_{\rm TLS} \approx \SI{374}{\kilo\hertz}$, similarly to previous works.\cite{Schloer2019,gertler2021protecting, levine2024demonstrating}

\begin{figure}
    \centering
    \includegraphics{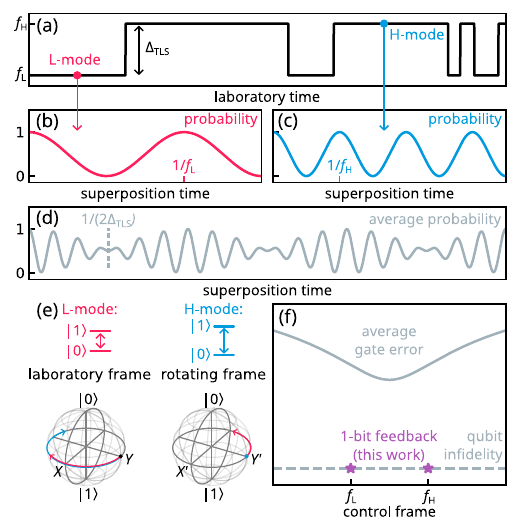}
    \caption{Dephasing and gate error of a bistable qubit. 
    (a) Illustration of a qubit switching sporadically between two modes, L and H, with frequencies $f_{\text{L}}$ and $f_{\text{H}}$, respectively.
    (b-c) Evolution of an equatorial superposition in the L and H modes.
    (d) Averaging over the two modes produces Ramsey beating with half-period
    $1/(2\Delta_{\rm TLS})$.
    (e) Bloch-sphere view of the relative phase accumulation in the laboratory
    and H-mode rotating frames.
    (f) The average gate error on a logarithmic scale (solid gray line, see main text), as a function of the qubit control frame, is expected to be dominating over the intrinsic qubit infidelity (gray dashed line). In contrast, our 1-bit feedback protocol allows the correct choice of the prevailing control frame [cf. Figures~\ref{fig:2}(a) and \ref{fig:3}(b)], resulting in a suppression of the average gate error (purple stars).
    }
    \label{fig:1}
\end{figure}

Conceptually, a bistable (but otherwise perfect) qubit randomly switches between two modes, as illustrated in Fig.~\ref{fig:1}(a). Without knowing its mode, such a qubit cannot be used reliably for quantum computations, as coherent superposition states would dephase with an average rate proportional to $\Delta_{\rm TLS}$. Our goal is to suppress this dephasing by mitigating TLS-induced telegraphic qubit-frequency fluctuations. In addition, bistability precludes high-fidelity gate operations, since the qubit frequency must be known before applying the correct physical control pulses.

The dephasing can be understood by preparing an equal superposition state and letting it evolve in the laboratory frame. If the qubit remained in the L mode, the probability to find the same state would oscillate sinusoidally with frequency $\fL$, as in Fig.~\ref{fig:1}(b). If it remained in the H mode, the probability would oscillate faster, as in Fig.~\ref{fig:1}(c). For a bistable qubit, however, the naive controller is ignorant of the mode, and the expected probability becomes an average of the two evolutions. As illustrated in Fig.~\ref{fig:1}(d), averaging over L and H modes with approximately equal weight yields a beating pattern, with characteristic half period $1/(2\Delta_{\rm TLS})$, which plays a central role in the mitigation protocol discussed below.

Figure~\ref{fig:1}(e) visualizes the dephasing process induced by the two modes in the XY plane of the Bloch sphere. In the laboratory frame, a superposition state (black dot) in the H mode evolves faster along the equator (blue curve) than in the L mode (red curve), thereby acquiring an additional phase shift relative to the L mode. Alternatively, in the rotating frame of the H mode---which we will adopt in Figure 2 for convenience---the state remains aligned along $Y'$, while in the L mode it acquires an equal but opposite phase shift.

Generally, the bistability precludes naive, high-fidelity operation of the qubit for several reasons. First, in order to accurately prepare a superposition state, for example by applying a $\frac{\pi}{2}$ pulse to $\ket{0}$, physical control pulses need to be applied to the qubit that typically depend on the qubit frequency. Second, execution of phase gates (Z rotations) typically also rely on knowing the qubit frequency accurately.  

Specifically, let's consider a $X_\pi$ pulse, which ideally rotates $\ket{0}$ onto $\ket{1}$ and vice versa, implemented by a standard Rabi driving process. We define the associated gate error as the state-preparation infidelity, $1-F = 1-\langle P(1)\rangle$, where $P(1)$ is the excited-state population after an $X_\pi$ pulse starting from $\ket{0}$, computed using the Rabi transition probability [Eq.~(\ref{eq:Rabi_prob}) of Appendix~\ref{app:infidelity}]. For a bistable qubit that is equally likely to be found in the L and H mode, the lowest average gate error is achieved by setting the control frame halfway in between $\fL$ and $\fH$, see solid gray line in Fig.~\ref{fig:1}(f).\footnote{computed from Eq.~(\ref{eq:avg_Rabi_prob}) of Appendix~\ref{app:infidelity} using $P(\fH)=P(\fL) = 1/2$} 
The same is true for the $Z$-rotation infidelity (see Appendix~\ref{app:z_fidelity}). In any case, a fixed choice for the control frame yields a gate error that \emph{on average} is much larger than the intrinsic qubit infidelity that may be achievable if the qubit was known to be in one specific mode (gray dashed line).    

The 1-bit feedback protocol below allows the controller to dynamically estimate the prevailing mode using only one single-shot measurement and to update the control frame accordingly, thereby recovering near-optimal gate fidelity. The remaining error (purple stars) is ultimately limited by the intrinsic qubit infidelity and the estimation error, detailed in Appendix~\ref{app:decoherence}, which takes into account all error sources unrelated to the average control-frame miscalibration.

\section{Protocol}
Our 1-bit feedback protocol is based on constructing a TLS-syndrome circuit that can discriminate the two qubit modes based on one single-shot readout outcome. 
The TLS-syndrome circuit evolves a state on the Bloch sphere equator for an optimal superposition time $\tau_{\text{opt}}$, such that the resulting state in the L mode becomes exactly antiparallel to the resulting state in the H mode. The two possible measurement outcomes $\{0,1\}$ in a suitable measurement basis can then exclusively identify the two qubit modes.\cite{luthi2018evolution} 
 
\begin{figure}
\includegraphics{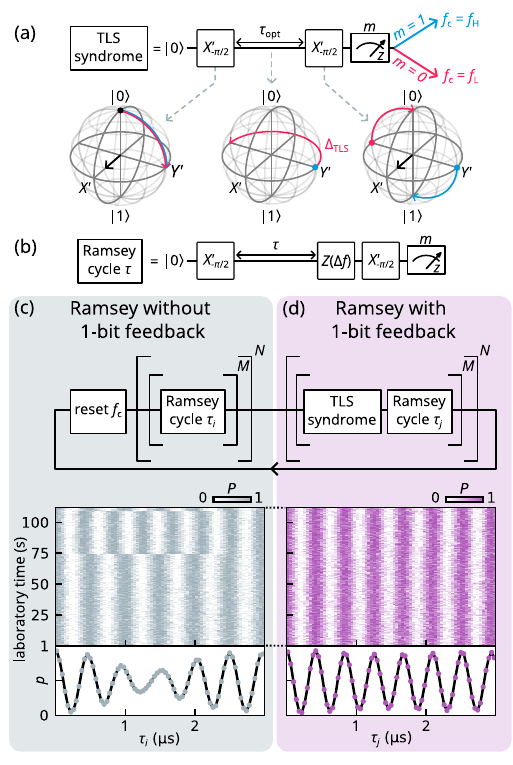}%
\caption{Mitigating dephasing of a bistable qubit by a feedback-controlled rotating frame.
    (a) In the two-level-system (TLS) syndrome cycle, the Bloch vector is initialized on the equator using an $X'_{-\pi/2}$ pulse. After precessing for a time $\tau_\text{opt}$, the L-mode state becomes antiparallel to the H-mode state (red and blue dot, respectively). A second $X'_{-\pi/2}$ pulse is then applied, followed by a measurement $m$ in the $Z$ basis, from which the syndrome (i.e., the qubit mode) is inferred and the control frame $\fc$ is updated accordingly.
    (b) In the Ramsey cycle, the superposition time $\tau$ is a stepped parameter, and the second $\frac{-\pi}{2}$ pulse is applied about an axis in the equatorial plane rotated by an angle $2\pi \Delta f \tau$ relative to the initial rotation axis. 
    (c–d) One repetition of the mitigation protocol traverses the loop (black arrow) exactly $M\times N$ times, yielding one row of the two two-dimensional plots below. $M=75$ sets the number of linearly spaced times $\tau_i$. $N=10$ sets the number of repetitions for each $\tau_i$ value, from which the fraction of $m=1$ outcomes, $P$, is calculated. Interleaving Ramsey cycles without feedback ($\fc$ set to the offline-calibrated value) with Ramsey cycles with feedback ($\fc$ set according to a preceding TLS syndrome measurement) allows direct comparison to validate the effectiveness of the 1-bit feedback.  
    (c) Each row plots $P(\tau_i)$ extracted from one repetition of the mitigation protocol, considering only the $75\times 10$ Ramsey cycles \emph{without} feedback.  
    (d) Each row plots $P(\tau_j)$ extracted from one repetition of the mitigation protocol, considering only the $75\times 10$ Ramsey cycles \emph{with} feedback. 
    In (c) and (d), repeating the mitigation protocol generates two-dimensional plots spanning nearly two minutes in laboratory time.  
    Inspection of the laboratory-time averages of $P(\tau_i)$ (purple and gray dots) clearly shows that the TLS-induced beating in the Ramsey fringes is suppressed by the 1-bit feedback protocol.
            }
        \label{fig:2}
\end{figure}

Specifically, the TLS syndrome cycle is defined by the circuit in Fig.~\ref{fig:2}(a) and implemented on the low-latency controller of our transmon qubit. In the rotating frame, the qubit Hamiltonian is
\begin{equation}
    \frac{\mathcal{H}(t)}{h}=-\dfrac{\Delta f -\Delta_{\rm TLS} \xi(t)}{2}\sigma_z,
\end{equation}
where $\sigma_{z}$ is the Pauli-$Z$ matrix, $\Delta f=\fc-\fq$ is the frequency detuning between the qubit frequency $\fq = \fH$ and the chosen rotating frame frequency $\fc$, $\Delta_{\rm TLS} = \fH - \fL$ as above, and $\xi(t)\in\{0,1\}$ represents the telegraphic time-dependent shift of the qubit frequency due to the TLS.

As illustrated by the Bloch spheres in Fig.~\ref{fig:2}(a), the TLS syndrome cycle first prepares a superposition state $\ket{\psi} = \left( \ket{0} + i \ket{1} \right)/\sqrt{2}$ using a $\frac{-\pi}{2}$ rotation around the $\hat{X}'$-axis. This $X'_{-\pi/2}$ pulse, which is calibrated for the H mode,  brings the state close to the equator also in the L mode, because the mode splitting $\Delta_{\rm TLS}$ is assumed to be small compared to the Rabi rate $\Omega$ of the $X'_{-\pi/2}$ pulse. While evolving as a superposition for time $\tau_{\text{opt}}$, the qubit state acquires a phase $
\phi = 2\pi \int_0^{\tau_{\text{opt}}} \delta_\text{q}(t') \, \mathrm{d}t'$, where $\delta_\text{q}(t) = \Delta f - \Delta_{\rm TLS} \xi(t)$. 
We assume $\xi(t)$ to be constant on the scale of the TLS syndrome cycle, resulting in $\phi = 2\pi[ \Delta f - \Delta_{\rm TLS} \xi(t)]\tau_\text{opt}$. (In the following, we drop the time dependence of $\xi(t)$ for ease of notation.) 
A second $X'_{-\pi/2}$ pulse then converts the accumulated phase into qubit polarization along $\hat{Z}$ , and the state of the qubit in the $Z$ measurement basis is acquired using standard dispersive readout techniques.

Ignoring the intrinsic qubit decoherence of each mode, state preparation and readout errors (see Appendix~\ref{app:decoherence} for further details on how they impact the estimation performance),  the probability of measuring an outcome $m\in \{0,1\}$ corresponding to the states $|0\rangle$ and $|1\rangle$ is given by
\begin{equation}\label{eq:ramsey}
    P(m)=\frac{1}{2}\{1+(-1)^{m+1}{\cos{[2\pi(\Delta f-\Delta_{\rm TLS}\xi)\tau_{\text{opt}}]}}\},
\end{equation}
where the binary variable $\xi$  (and thus the qubit mode) is what we want the controller to estimate. 

The optimal probing time is determined by requiring that the two possible qubit frequencies produce maximally distinguishable states. Without loss of generality, in the rotating frame of the H mode ($\fc=\fH$), the H mode acquires no phase during the superposition time (blue dot in the center Bloch sphere), while the L mode accumulates a phase at a rate set by the frequency difference $\Delta_{\rm TLS}$ (red solid line). After a superposition time $\tau$, the relative phase between the two states is therefore $2\pi \Delta_{\rm TLS}\tau$. Maximum discrimination is achieved when this phase difference equals $\pi$, placing the two Bloch vectors at opposite points on the equator just before the second $X'_{-\pi/2}$ pulse.  
In the limit $\Delta_{\rm TLS} T_2 \gg 1$ (where $T_2\approx\SI{43}{\micro\second}$ is the mode-intrinsic dephasing time, see Appendix~\ref{app:decoherence}), this condition yields the optimal probing time
\begin{equation}
\tau_{\text{opt}} \approx \frac{1}{2\Delta_{\rm TLS}},
\end{equation}
which maximizes sensitivity to the TLS-induced frequency shift and yields $\tau_{\text{opt}} \approx \SI{1.33}{\micro\second}$ for $\Delta_{\rm TLS} \approx \SI{374}{\kilo\hertz}$.

In our experiment, the assumption of negligible decoherence during the TLS syndrome cycle is justified because the total qubit cycle duration is short compared to the decoherence timescales of tens to $\approx\SI{100}{\micro\second}$, detailed in the randomized benchmarking experiment below. Moreover, the TLS switching time of tens of seconds is much longer than one TLS syndrome experiment. Specifically, gate pulses are of order $\SI{48}{\nano\second}$, the superposition time is $\tau_{\text{opt}} \approx \SI{1.33}{\micro\second}$, the qubit readout takes $\approx\SI{2}{\micro\second}$, and the resonator reset time, used to deplete residual photons after the readout pulse, is $\approx\SI{6}{\micro\second}$.

In summary, since the qubit frequency $f_\text{q}$ only assumes two known discrete values, the mode of the bistable qubit can be inferred from a single measurement outcome $m$, rather than from several measurements.\cite{Arshad2024, Berritta2024b, liu2024observation, Berritta2025_FBS, hutin2025preparing} At the end of the TLS syndrome cycle, the control frame $\fc$ is updated accordingly. Each single-shot measurement thus provides one bit of information about the TLS state: By storing this information in the controller and feeding it back in real time, we show experimentally below that the uncertainty of the qubit frequency is suppressed, thereby mitigating telegraphic fluctuations of the bistable qubit.

\section{Results}
Effectiveness of the one-shot feedback is validated by estimating the qubit mode and adjusting $\fc$ in real time, thereby suppressing TLS-induced Ramsey beating of the transmon qubit. A standard Ramsey experiment consists of $M$ Ramsey cycles, in which the superposition time $\tau$ is linearly incremented. Repeating such an experiment $N=10$ times for each choice of $\tau$ allows us to calculate the fraction of $m=1$ outcomes, $P$. The average readout fidelity is 94\% and its impact on the estimation scheme is discussed in Appendix~\ref{app:decoherence}. 
Within each Ramsey cycle, defined by the circuit in Fig.~\ref{fig:2}(b), the evolution of superposition states is followed by a virtual $Z$ gate, which sets the phase $\Delta f \tau$ of the return probability function [Eq.~(\ref{eq:ramsey})].\cite{McKay2017} Here, ``virtual'' means that the phase offset is applied by shifting the phase of the subsequent $X'_{-\pi/2}$ pulse, which projects the Bloch vector back onto the $\hat{Z}$ axis. 
The intentional detuning introduced via the virtual $Z$ gate sets the period of the resulting Ramsey fringes, yielding a high-visibility oscillatory behavior that is easy to fit. 

To validate the single-shot syndrome extraction, we interleave TLS syndrome cycles with Ramsey cycles, half of which do not make use of the mode estimation and half of which do, as shown schematically in Figs.~\ref{fig:2}(c)–(d). 
\footnote{For the virtual detunings, we choose $\Delta f = \SI{2}{\mega\hertz}$ for Ramsey cycles without mode estimation and $\Delta f = \SI{2.33}{\mega\hertz}$ for Ramsey cycles with mode estimation.}  
Around $\SI{75}{\second}$ and $\SI{109}{\second}$ of laboratory time, the qubit frequency sporadically switches from the H mode to the L mode and vice versa. 
The suppression of TLS-induced beating in the lower panels of Figs.~\ref{fig:2}(c)–(d) demonstrates that the estimated qubit mode obtained from the TLS syndrome measurement provides accurate knowledge of the fluctuating parameter $\fq$.

Remarkably, our approach requires only one single-shot measurement outcome, compared to the tens or hundreds of binary measurements in nonadaptive protocols.\cite{Gebhart2023, Park2025}  
This is considerably less than even the handful of measurements required in the very efficient binary search protocol by \citet{Berritta2025_FBS}, which adaptively mitigated $1/f$ noise. 
Since here, the qubit ``drift'' is limited to a bistability between two known frequencies, only one single-shot measurement is required. As a consequence, the bandwidth of the method is $\approx\SI{136}{\kilo\hertz}$ (set by the readout time, resonator cooldown, and Ramsey superposition time), which is more than one order of magnitude higher than the above-mentioned works. This high bandwidth is enabled by the two-state telegraphic structure of the bistable qubit frequency, for which a single optimally timed measurement discriminates the active modes. Relative to earlier single-shot charge-parity detection,\cite{Rist2013} the contribution demonstrated here is the use of the measurement outcome for subsequent gate control and the resulting improvement in gate fidelity. In other words, in this work the controller does not continuously track a drifting parameter, but it performs hypothesis testing between two discrete Hamiltonians.
\begin{figure}
\includegraphics{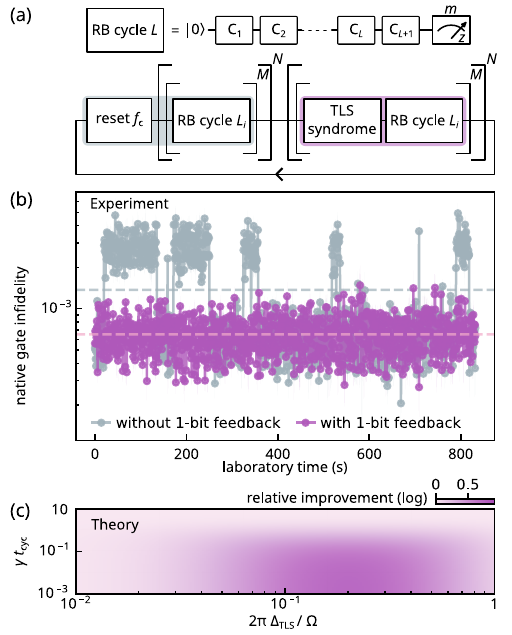}
\caption{Quantifying improvement of gate fidelities by randomized benchmarking (RB).
			(a) Upper schematic: RB cycle, where the circuit depth $L$ is a stepped parameter. Lower schematic: Modified RB protocol, interleaving $N=100$ random gate sequences of length $L_i$ stepped logarithmically $M = 12$ times, with and without estimating the qubit mode from the TLS syndrome. Fitting the averaged exponential decay with increasing $L_i$, yields native gate infidelities for the traditional gates (without feedback) and the stabilized gates (with feedback). 
			(b) Fitted native gate infidelities, obtained by running the RB protocol over a period of 14 minutes, reveal two-level fluctuations of the infidelity for sequences without feedback (gray dots) and consistently low infidelities with feedback (purple dots). Dashed lines are the averaged gate infidelities.
            (c) Theoretical relative improvement of the state-preparation infidelity with and without estimation as a function of the normalized TLS splitting $\Delta_{\rm TLS}$ and switching rate $\gamma$ (see main text). 
		}
		\label{fig:3}
\end{figure}

Finally, we quantify the improvement of single-qubit gate fidelities attributable to the 1-bit feedback by randomized benchmarking (RB).\cite{Knill2008} RB applies $N$ random sequences of $L$ Clifford gates to an initial state, followed by an $(L+1)$th ``recovery'' gate chosen such that each sequence ideally implements the identity operation in the absence of errors [see upper schematic of Fig.~\ref{fig:3}(a)]. Averaging the single-shot measurement outcomes over many random sequences reduces the resulting error structure to a single-exponential decay as a function of $L$, from which the average gate infidelity is extracted via a least-squares fit.

Our modified RB protocol is shown in the lower schematic of Figure~\ref{fig:3}(a). For each RB estimate without feedback [one gray point in panel (b)], the controller first sets $\fc = \fH$ and performs $N=100$ random sequences, each evaluated at $M=12$ logarithmically spaced depths $L_i$ (powers of 2 up to 2048). Subsequently, for each RB estimate with feedback [one purple point in panel (b)], the controller again performs $N=100$ random sequences evaluated at the same $M=12$ depths, but now $\fc$ is updated based on a TLS syndrome measurement acquired before \textit{each} RB cycle at depth $L_i$.\footnote{Thus, each purple point in Fig.~\ref{fig:3}(b) involves $N\times M = 100\times 12 = 1200$ TLS-syndrome outcomes, which enter the fitted RB decay as an averaged error channel rather than as isolated events in the plotted gate-infidelity trace.} 

This interleaving of RB repetitions with and without 1-bit feedback allows us to isolate the effects of the feedback. 
The controller continuously runs the RB protocol for approximately 14 minutes, allowing us to observe telegraphic fluctuations of the fitted gate infidelity for sequences without feedback (gray dots), while consistently low infidelities are observed for the stabilized sequences (purple dots).  The average native gate infidelity (mean $\pm$ standard deviation) is $(1.4 \pm 1.1) \times 10^{-3}$ without feedback and $(0.65 \pm 0.18) \times 10^{-3}$ with feedback [dashed lines in Fig.~\ref{fig:3}(b)]. The shaded regions, barely visible due to the small uncertainty, show the 68\% confidence interval for the fitted gate infidelities. \footnote{One data point in the no-feedback trace (laboratory time $\SI{575}{\second}$) was excluded from Fig.~\ref{fig:3}(b), as the corresponding randomized-benchmarking fit failed} The main result here is that the average gate infidelity with feedback is significantly lower than the average infidelity without feedback.

Importantly, the 1-bit feedback protocol does not introduce any additional errors in comparison with the baseline of the reference experiment (no feedback). Quantitatively, the measured single-qubit gate infidelities for the 1-bit feedback protocol are slightly higher than the theoretical decoherence limit, approximated by $t_{\text{gate}}(1/T_1 + 1/T_{\phi})/3 \approx 5 \times 10^{-4}$, using experimental values for gate time $t_{\text{gate}} \approx \SI{48}{\nano\second}$, qubit relaxation time $T_1 \approx \SI{74}{\micro\second}$, and qubit pure dephasing time $T_{\phi} \approx \SI{61}{\micro\second}$.\cite{Malley2015} The feedback protocol outperforms the no-feedback case, suppressing the largest observed errors from $2.8 \times 10^{-3}$ down to approximately $0.65 \times 10^{-3}$, corresponding to an improvement of about 77\% due to the control stabilization.

Figure~\ref{fig:3}(c) shows the theoretically computed relative improvement of state-preparation infidelity with active control $(1-F_{\rm active})$ over blind driving $(1-F_{\rm blind})$, estimated in Appendix~\ref{app:rtn_error}. The TLS splitting $\Delta_{\rm TLS}$ is normalized with respect to the Rabi frequency $\Omega$, while the TLS switching rate $\gamma$ is expressed in terms of the dimensionless product $\gamma t_{\rm cyc}$, where $t_{\rm cyc}$ is the total estimation cycle time. 
This theoretical improvement is plotted for the asymmetric drive condition ($\fc = \fH$) used in the experiment. When the switching rate becomes comparable to the cycle time ($\gamma t_{\rm cyc} \gtrsim 1$) the active estimates lag behind the switching dynamics, and the relative improvement saturates at the static blind-hardware baseline. As $\Delta_{\rm TLS}$ decreases, the optimal discrimination time $t_{\rm opt}$ increases and approaches the intrinsic coherence time of the qubit, thereby reducing the protocol performance. For the symmetric case, $\fc = (\fL+\fH)/2$, wrong estimates can instead lead to a negative relative improvement, see Appendices~\ref{app:cons_est_error} and~\ref{app:rtn_error}.
\section{Conclusions}
 
This work presents an experimental demonstration of operating a bistable qubit with high fidelity using a low-latency FPGA-based controller and just one recent single-shot measurement outcome. The 1-bit feedback scheme is essential for enabling control pulses to compensate for discrete qubit-frequency shifts caused by a strongly coupled TLS, suppressing TLS-induced Ramsey beating and reducing gate infidelities by up to 77\%. The protocol assumes that the TLS switching rate is much slower than the estimation bandwidth ($\approx\SI{136}{\kilo\hertz}$) and that the qubit modes separation exceeds the qubit decoherence rate. 

Current efforts to reduce the detrimental impact of TLSs in solid-state qubits may lead to an era of next-generation quantum processors that suffer from just a few imperfect qubits with sporadic instabilities. Efficient active mitigation of discrete instabilities may therefore become a key resource for operating such processing units to their fullest capabilities (after all, each additional qubit increases the associated Hilbert space by a factor of two). It has been shown that TLS frequencies drift over repeated cooldowns, whereas the overall number of TLSs does not.\cite{Shalibo2010, Zanuz2024} Because individual TLSs can induce non-Markovian, telegraphic parameter fluctuations, real-time frequency-tracking protocols may serve as practical tools to suppress these fluctuations and minimize temporal correlations of the noise.\cite{Park2025,Berritta2025_FBS} This is particularly relevant for ancilla qubits, as the absence of such correlations is a fundamental assumption for the syndrome extraction measurements required in fault-tolerant quantum error correction.

While improvements in design, materials and fabrication are ongoing and worth pursuing, our feedback protocol provides a complementary, control-based approach for real-time calibration of TLS-affected qubits at the information limit set by the qubit's intrinsic entropy, thereby increasing the computational volume of near-perfect quantum processing units at comparatively low cost. 


%
%

%


\begin{acknowledgments}
F.B. acknowledges funding from the European Union's Horizon Europe research and innovation programme under grant agreement 101204890 (HORIZON-MSCA-2024-PF-01). 
J.A.K. acknowledges support from the Dutch National Growth Fund (NGF) as part of the Quantum Delta NL programme.
J.D. acknowledges funding from the Research Council of Norway through Grant No. 353894.
F.K. acknowledges support by the Army Research Office under Award Number W911NF-24-2-0043 and the EU under ERC grant agreement No. 856526 and through the H2024 QLSI2 initiative under project No. 101174557. 
Any opinions, findings, conclusions, or recommendations expressed in this material are those of the authors and do not necessarily reflect the views of the Army Research Office, the US Government, the European Union, or the European Research Council. Neither the European Union nor the granting authority can be held responsible for them.
\end{acknowledgments}

\section*{Author declarations}
\subsection*{Conflict of interest}
The authors have no conflicts to disclose.

\subsection*{Author contributions}
\textbf{Fabrizio~Berritta}: Funding acquisition (lead); Conceptualization (equal); Methodology (lead); Investigation (lead); Formal analysis (lead); Visualization (lead); Writing – original draft (lead); Writing – review and editing (equal).
\textbf{Jan~A.~Krzywda}: Methodology (supporting); Formal analysis (supporting); Writing – original draft (supporting); Writing – review and editing (equal).
\textbf{Tom~Dvir}: Investigation (supporting); Software (supporting); Writing – review and editing (equal).
\textbf{Paul~Buttles}: Resources (lead); Writing – review and editing (equal).
\textbf{Stanislav~Eilhart}: Resources (lead); Writing – review and editing (equal).
\textbf{Jeroen~Danon}: Methodology (supporting); Formal analysis (supporting); Writing – review and editing (equal).
\textbf{Ferdinand~Kuemmeth}: Conceptualization; Investigation; Supervision (lead);  Writing – review and editing (equal).

\section*{Data availability}
The data that support the findings of this study are openly
available in Zenodo at
\url{https://doi.org/10.5281/zenodo.23424770}.\cite{Berritta2026Data}

\appendix
\section{Setup and device}\label{app:setup_device}
\subsection{Experimental setup}
We use a superconducting qubit array operated at the mixing chamber stage of a Bluefors XLD1000 dilution refrigerator with a base temperature below $\SI{10}{\milli\kelvin}$ at the Israeli Quantum Computing Center (IQCC). We implement the estimation protocol on one of the available transmon qubits, which comprises a DC SQUID with asymmetric junctions. A commercial FPGA-powered controller (Quantum Machines OPX1000) applies high-frequency waveforms for qubit control and single-shot readout, using on-chip Z and XY control lines and readout resonators respectively.\cite{Krantz2019, Blais2021}. The OPX1000 includes real-time classical processing with fast analog feedback programmed in QUA software. The readout tone and control lines are attenuated in the cryostat to remove excess thermal photons from higher-temperature stages, and filtered at the mixing chamber. The Z control lines are passively attenuated by $\SI{10}{\decibel}$ at room temperature and $\SI{20}{\decibel}$ inside the cryostat, then filtered using a Quantum Microwave QMC-CRYOIRF-001 low cut-off IR filter. The XY control lines and readout tone, approximately $\SI{6.3}{\giga\hertz}$, are passively attenuated by $\SI{60}{\decibel}$ inside the cryostat and filtered with an RLC Electronics F-30-8000-R low pass filter and Quantum Microwave QMC-CRYOIRF-003 high cut-off IR filter. The transmitted signal from the feedline goes through a high-cutoff IR filter, a Keenlion 4-8~$\SI{}{\giga\hertz}$ bandpass filter, and Low Noise Factory 4-8~$\SI{}{\giga\hertz}$ single and double junction isolators to prevent back reflections from the amplification chain. A high-electron mobility transistor amplifier (LNF-LNA-4-8G) thermally anchored at the $\SI{3}{\kelvin}$ stage amplifies the readout signal. At room temperature, the readout line is again amplified (Narda-Miteq LNA-40-04000800-07-10P). The device sample is aluminum wirebonded into a custom-made sample mount using a high density of package-to-chip and chip-to-chip wirebonds to ensure proper grounding. The sample mount comprises a printed circuit board and superconducting aluminum enclosure designed to suppress stray microwave and infrared photons. The sample mount is placed inside a light-tight cryoperm can to further reject stray photons and provide magnetic shielding. Both the sample mount and the magnetic shield can are supplied by Qolab. The tunable transition frequency of our transmon is controlled by an external magnetic flux $\Phi_{\text{ext}}$ applied via mutual coupling to the $Z$ line. We bias the qubit to the maximum frequency where it is first-order insensitive to flux noise.
\subsection{Device fabrication}

The transmons were fabricated on a high-resistivity Si wafer ($>\SI{10}{k\Omega-\text{cm}}$) cleaned with diluted HF to remove native oxide prior to sputter deposition of an aluminum base layer. The base layer was defined on a positive photoresist layer written using a Heidelberg DWL 66+ laser writer, then developed and etched using a TMAH-based developer. Dolan bridge Josephson junctions were fabricated using electron beam lithography (EBL), including a bilayer stack of EBL resist exposed on an Elionix 100 keV electron beam writer and developed with a dilute solvent mixture. The counterelectrode was deposited using an electron beam evaporator. Liftoff was performed in an NMP-based solvent, then the sample was sonicated in a solvent bath before being diced and packaged.
\section{Impact of bistability on qubit operations}
In this section, we discuss how the bistable nature of the qubit affects state preparation, coherent operations, and readout fidelity.

\subsection{Driven X-gate fidelity without active estimation}\label{app:infidelity}
As a measure of qubit operation fidelity, we use the fidelity of an $X_{\pi}$ pulse applied to the qubit in the ground state.
First, we consider the fidelity $F_X = \langle P(1)\rangle$ of the final state for the case without active feedback (``blind'' drive). We assume that the qubit is initialized in $\ket{0}$ and then driven at a frequency $\fc$ with a Rabi amplitude $\Omega$ for a time $t_\pi = \pi/\Omega$, which implements an X-gate at resonance, i.e., when $\fc$ matches the qubit transition frequency. However, the qubit frequency stochastically switches between $\fL$ and $\fH$ due to the TLS. The average fidelity of an X-gate applied in this way is thus
\begin{equation}\label{eq:avg_Rabi_prob}
    \bar F_X(\fc) = P(\fL)F_X(\fc|\fL) + P(\fH) F_X(\fc|\fH),
\end{equation}
where $F_X(\fc|f_i)$ is the fidelity given a drive detuned from $f_i$ by $\delta_i = 2\pi(\fc - f_i)$ and $P(f_i)$ is the fraction of the time the qubit spends in state $i$. Using the standard Rabi formula we find the fidelity
\begin{equation}\label{eq:Rabi_prob}
    F_X(\fc|f_i) = \frac{\Omega^2}{\Omega^2 + \delta_i^2} \sin^2\left(\frac{\sqrt{\Omega^2 + \delta_i^2} }{2\Omega}\pi\right).
\end{equation}

In the weak-noise limit, where the detunings $\Delta_{\rm H,L} \ll \Omega$, we approximate
\begin{equation}\label{eq:FXquad}
    F_X(\fc|f_i) \approx 1 - \frac{\delta_i^2}{\Omega^2}.
\end{equation}
From maximizing the average fidelity $\bar F_X$ we find an optimal drive frequency $\fc^{\text{opt}} = P(\fL) \fL + P(\fH) \fH$ at the weighted average of the two qubit frequencies. Substituting this back gives as maximum average fidelity
\begin{equation}
   \bar F_X(\fc^\text{opt})  = 1 - \left(\frac{2\pi\Delta_{\rm TLS}}{\Omega}\right)^2 P(\fL) P(\fH) .
\end{equation}
This indicates that the infidelity scales to leading order with the square of the splitting $\Delta_{\rm TLS}$. Importantly, the infidelity is \textit{maximized} (worst-case scenario) when the TLS populations are equal, $P(\fL) = P(\fH) = 1/2$. In this worst case, the detuning for each state is at the optimal point exactly $\Delta_{\rm TLS} / 2$, leading to a bound on the infidelity
\begin{equation}
    1-\bar F_X \leq \frac{\pi^2 \Delta_{\rm TLS}^2}{\Omega^2}.
\end{equation}

Note that the above derivation only considers the coherent error induced by the uncompensated detuning. To leading order, the total average gate error is the sum of this coherent error and the intrinsic error limit imposed by qubit decoherence during the pulse and SPAM, i.e.,
\begin{equation}
\label{eq:blind_fidelity}
    1-F_\text{blind} \approx (1 - \alpha e^{-t_\pi/T_2}) + \alpha'\frac{\pi^2 \Delta_{\text{TLS}}^2}{\Omega^2}.
\end{equation}
Here $\alpha'$ accounts for the reduction in visibility due to SPAM errors and decoherence, as defined in Appendix~\ref{app:decoherence}.

We finally briefly address the opposite limit of large frequency splitting, $\Delta_{\rm TLS} \gtrsim \Omega$. In this regime, the quadratic approximation of Eq.~(\ref{eq:FXquad}) breaks down. The optimal drive frequency no longer lies at the weighted average, as the fidelity function becomes bimodal with local maxima at $\fL$ and $\fH$. Driving at the average frequency results in a detuning $\Delta \approx \Delta_{\rm TLS}/2$ that exceeds the power-broadened linewidth, causing the Rabi amplitude to vanish. 
In this case, the optimal blind strategy transitions to targeting the most probable frequency mode (e.g., $\fc = \fL$), accepting the resulting loss of fidelity if the qubit is in the other state.
For an active strategy in this regime, we propose a feedback strategy that uses a modified syndrome cycle: The qubit is assumed to be in the most probable mode, and exposed to $\pi$ rotations (calibrated for that mode) that flip $\ket{0}$ to $\ket{1}$ and vice versa only in that mode. Measurement outcomes confirming flipping then confirm the mode, whereas measurement outcomes confirming the absence of flipping would indicate that the qubit is in the other mode.  

\subsection{Driven X-gate fidelity with active estimation}\label{app:cons_est_error}
We now analyze the fidelity improvement gained by active estimation. For simplicity let's assume $P(f_L) = P(f_H) = 0.5$, and denote probability of correctly estimating the qubit frequency as $1-p_\text{err}$. 
The average control fidelity is the weighted sum of the fidelities for the correct and wrong estimation cases:
\begin{equation}
    F_{\text{active}} = (1-p_\text{err}) F_{\text{correct}} + p_\text{err} F_{\text{wrong}}.
\end{equation}
The fidelity in the correct case, $F_{\text{correct}}= \alpha \exp(-t_\pi/ T_2)$, is limited only by intrinsic decoherence and SPAM. 
In the wrong case, the pulse suffers from \textit{both} the intrinsic decoherence and the unitary error due to the detuning $\Delta_{\rm TLS}$. To leading order, these errors multiply:
\begin{equation}
    F_{\text{wrong}} \approx F_{\text{correct}} \left( 1 - \frac{4\pi^2 \Delta_{\rm TLS}^2}{\Omega^2} \right).
\end{equation}
Substituting this back into the average fidelity equation we find to leading order for the infidelity of the X-gate under active estimation
\begin{align}
\label{eq:active_fidelity}
    1-F_{\text{active}} \approx (1 - \alpha')  + \alpha' p_\text{err}  \frac{4\pi^2 \Delta_{\rm TLS}^2}{\Omega^2}.
\end{align}

To assess the benefits of the active estimation strategy, we must compare Eq.~(\ref{eq:active_fidelity}) against the optimal blind pulse fidelity. Crucially, the limit of this blind baseline depends strictly on the drive strategy.

In a symmetric strategy, the drive frequency is parked at the average frequency $f_c = (f_L+f_H)/2$. Both states experience a detuning of $|\Delta| = \Delta_{\rm TLS}/2$, yielding a blind infidelity of $1-F_{\rm blind}^{\rm sym} \approx \alpha'(\pi\Delta_{\rm TLS}/\Omega)^2$. Comparing this to Eq.~(\ref{eq:active_fidelity}), active estimation outperforms the blind average-frequency drive as long as the estimation error is $p_\text{err} < 1/4$.

Conversely, in the asymmetric strategy the hardware is locked to the high-frequency mode ($f_c = f_H$). If active feedback is disabled, the blind pulse remains perfectly resonant with $f_H$ (zero unitary error), but misses $f_L$ by the full splitting $\Delta_{\rm TLS}$. Averaging these outcomes yields a blind infidelity of $1-F_{\rm blind}^{\rm asym} \approx \frac{\alpha'}{2}(0) + \frac{\alpha'}{2}(4\pi^2\Delta_{\rm TLS}^2/\Omega^2) = 2\alpha'(\pi\Delta_{\rm TLS}/\Omega)^2$. Because missing one mode entirely is quadratically worse than missing both by half, the asymmetric blind baseline is twice as erroneous. In the main text, we use the asymmetric drive because it makes the telegraphic switching of the gate infidelity directly visible. With the conventional symmetric drive, both qubit modes would experience equal-magnitude detuning, so the no-feedback gate infidelity would be expected to remain approximately constant as the TLS switches between modes. The active protocol on this architecture thus provides a net benefit as long as $p_{\text{err}} < 1/2$. In Appendix~\ref{app:decoherence} we focus on computing the estimation error in the presence of decoherence.

\subsection{Virtual Z-gate fidelity}\label{app:z_fidelity}
Single-qubit $Z$ rotations are typically implemented as virtual gates by updating the reference phase of the microwave drive in software. While this operation is instantaneous, an error in the frequency estimation leads to an incorrect frame rotation update. Because the software frame and the physical qubit are evolving at different frequencies, separated by a detuning $\Delta$, a phase error $\delta \phi = \Delta t_g$ accumulates over the duration $t_g$ of any subsequent physical operation or idle period. This accumulated phase error, similarly to the X-gate error, is proportional to $\langle \delta \phi^2 \rangle = \langle \Delta^2 \rangle t_g^2 = (2\pi \Delta_{\rm TLS} t_g)^2/4 $.

In the blind operation regime, the software frame tracks the average frequency, resulting in a constant detuning magnitude $|\Delta| = \Delta_{\rm TLS} / 2$ for every qubit cycle. Conversely, with active feedforward, the frame tracks the estimated frequency. This results in zero error with probability $1-p_{\text{err}}$, and a large error corresponding to $\Delta = \Delta_{\rm TLS}$ with probability $p_{\text{err}}$. Consequently, the fidelity threshold derived above ($p_{\text{err}} < 0.25$ for symmetric, $p_{\text{err}} < 0.5$ for asymmetric) apply equally to Z-gates. However, unlike the blind strategy which yields a uniform error rate across all shots, active feedback produces a bimodal error distribution consisting of a majority of high-fidelity shots mixed with rare large-error outliers. This bimodal distribution can lead to non-exponential decays in randomized benchmarking experiments, as the sequence fidelity becomes a weighted sum of two distinct decay rates.

\section{Decoherence and SPAM}\label{app:decoherence}
In this section, we discuss how decoherence and State Preparation and Measurement (SPAM) errors affect the estimation performance of the bistable qubit.

Decoherence modifies the likelihood function of the Ramsey probing cycle [Eq.~(\ref{eq:ramsey})] by reducing the visibility of the oscillations. In the presence of decoherence and SPAM, the likelihood function becomes
\begin{equation}
    P(m|\xi,\tau) = \frac{1}{2} + (-1)^{m+1}\frac{\alpha}{2} e^{-\tau/T_2} \cos[2\pi(\Delta f - \xi \Delta_{\text{TLS}})\tau],
\end{equation}
where $T_2$ is the effective qubit coherence time and $\alpha \leq 1$ accounts for the reduction in visibility due to SPAM errors. The exponential factor suppresses the oscillation contrast, increasing the difficulty of discriminating between the two possible qubit frequencies.

We first determine the optimal superposition time $\tau_{\text{opt}}$ that maximizes the distinguishability between the two cases, $\xi=0$ and $\xi = 1$ defined as the signal contrast:
\begin{align}
S(\Delta_{\rm TLS}, \tau) &= |P(1| \xi = 0, \tau) - P(1|\xi= 1, \tau)| \nonumber \\ &=  \left|\alpha \,e^{-\tau/T_2} \sin[2\pi(\Delta f-\Delta_{\rm TLS}/2)\tau] \sin(\pi \Delta_{\rm TLS} \tau)\right|.
\end{align} 
When the measurement is taken in rotating frame with respect to the case $\xi=0$, we set $\Delta  f = 0$ and have ${S(\Delta_{\rm TLS}, \tau) = \alpha e^{-\tau/T_2} \sin^2(\pi \Delta_{\rm TLS} \tau)}$. Maximizing $S(\Delta_{\rm TLS}, \tau)$ with respect to $\tau$  yields 
\begin{equation}
\label{eq:transcendental}
    \tau_{\text{opt}} = \frac{1}{\pi \Delta_{\rm TLS}} \arccos\left( \frac{1}{\sqrt{1+4\pi^2 \Delta_{\rm TLS}^2 T_2^2}} \right).
\end{equation}
In the limit where the frequency splitting is large compared to the linewidth ($\Delta_{\rm TLS} T_2 \gg 1$), the optimal time is dominated by the oscillation frequency rather than the decay, approaching $\tau_\text{opt} \approx 1/(2\Delta_{\rm TLS})$.

SPAM errors reduce the visibility by a constant factor $\alpha$, which introduces a uniform reduction in distinguishability $S(\Delta_{\rm TLS}, \tau)$ for all $\tau$. Therefore, while SPAM errors do not shift the optimal superposition time $\tau_{\text{opt}}$, they degrade the overall estimation performance. The maximum distinguishability at $\tau_{\text{opt}}$ is given by
\begin{align}
S(\Delta_{\rm TLS}, \tau_{\text{opt}}) {} & {} = \alpha
\frac{4\pi^2 \Delta_{\rm TLS}^2 T_2^2}{1+ 4\pi^2 \Delta_{\rm TLS}^2 T_2^2}e^{-\frac{\arctan (2\pi\Delta_{\rm TLS}T_2) }{\pi \Delta_{\rm TLS} T_2}} \nonumber\\ {} & {} 
= 1 - 2p_\text{err},
    \label{eq:perr}
\end{align}
where $p_{\text{err}} = 0.5$ represents the limit of random guessing (zero contrast). For instance, achieving a correct discrimination probability of at least 95\% requires $1 - p_{\text{err}} \geq 0.95$, or $S \geq 0.9$. In this experiment, $p_{\text{err}} \approx 9\%$, which is significantly below the $50\%$ utility threshold of the asymmetric architecture defined in Appendix~\ref{app:cons_est_error}.

\section{Finite pulse duration in the Ramsey sequence}\label{app:finite_pulse}

The derivation above assumes instantaneous state preparation and measurement pulses. In a true Ramsey probing cycle, the initial and final $X_{\pi/2}$ pulses have a finite duration, which induces additional errors that depends on the drive frequency. Below we analyze two cases: The symmetric one, where the pulses are driven at the average frequency of the two states $\fc = (\fL + \fH)/2$, and the one used in the paper $\fc = \fH$. We show that the latter strategy can offer additional error cancellation.

\subsection{Symmetric drive}
In this frame, the detunings are symmetric: $\Delta_L = \pi\Delta_{\text{TLS}}$ and $\Delta_H = -\pi\Delta_{\text{TLS}}$. Because the detuning magnitudes are identical, both states experience the same generalized Rabi frequency, $\Omega' = \sqrt{\Omega^2 + (\pi\Delta_{\text{TLS}})^2}$, but rotation takes place around differently tilted axes. As a result when nominal duration $t = \pi/(2\Omega)$ is used, reduction in the Ramsey fringe visibility is a compounded effect of the error along the Z axis (amplitude error) and in the XY plane (phase error). One can show that the amplitude error is heavily suppressed, resulting in relatively small renormalization of the SPAM coefficient. This allows us to neglect the correction and concentrate on a state-dependent phase shift $\delta \phi_i \approx 
\pm \pi\Delta_\text{TLS}/\Omega$, which modifies the phase of the state vector in the XY-plane.

To resolve the sign of the symmetric detunings, the measurement pulse must project the orthogonal quadrature, requiring an $X'_{\pi/2} - \tau - Y'_{\pi/2}$ sequence. The phase accumulates twice, acting as an effective timing offset $\delta\tau = 2/\Omega$ added to the free evolution. Incorporating this phase offset yields the true distinguishability function:

\begin{equation} S_{\text{true}}(\Delta_{\text{TLS}}, \tau) = \alpha  e^{-\tau/T_2}\sin^2\left[\pi\Delta_{\text{TLS}}\left(\tau + \frac{2}{\Omega}\right)\right].
\end{equation}
Maximizing this function yields a shifted transcendental equation for the optimal superposition time:
\begin{equation}
    \tan\left[\pi\Delta_{\text{TLS}}\left(\tau_{\text{opt}} + \frac{2}{\Omega}\right)\right] = 2\pi\Delta_{\text{TLS}}T_2.
\end{equation}

\subsection{Asymmetric drive}
In the experimental configuration used in this work, the sequences are driven asymmetrically at $\fc = \fH$. In this rotating frame, the high-frequency state experiences no detuning ($\Delta_H = 0$). The low-frequency state, however, absorbs the full frequency splitting, $\Delta_L = 2\pi\Delta_{\text{TLS}}$, and consequently rotates around a heavily tilted axis during the pulses.

If a standard, non-inverted projection pulse $X'_{\pi/2}$ is used for readout, where prime denotes rotated frame, $\fL$ state accumulates a finite-pulse phase error of approximately $2\pi\Delta_{\text{TLS}}/\Omega$ during state preparation, and an identical error during measurement. Because these errors add constructively, the total phase shift yields exactly the same effective timing offset, $\delta\tau = 2/\Omega$, and the identical distinguishability function as derived for the symmetric case above.

\section{Relation between estimation bandwidth and estimation error}
\label{app:rtn_error}
To model the qubit coherence under a finite telegraph switching rate $\gamma$ and estimation bandwidth $\text{BW}$, one can use a framework originally used by Anderson and Kubo to describe spectral line shape narrowing (motional narrowing) in nuclear magnetic resonance due to random environmental fluctuations.\cite{anderson1954, kubo1954} In our case, the total transverse coherence $C(t) = \langle \sigma_+(t) \rangle$ of a qubit subject to a bistable frequency shift $\pm \pi \Delta_{\text{TLS}}$ satisfies the classical damped oscillator equation
\begin{equation}\label{eq:cdo}
    \ddot{C} + \gamma \dot{C} + \pi^2 \Delta_{\rm TLS}^2 C = 0,
\end{equation}
where $C(t)$ is expressed in the rotating frame of the average of the two frequencies.
Imposing $C(0)=1/2$, the general solution to Eq.~(\ref{eq:cdo}) reads in the underdamped regime ($\gamma < 2\pi\Delta_{\rm TLS}$) as
\begin{equation}\label{eq:solcdo}
    C(t) = e^{-\gamma t / 2} \left[ A e^{-i\beta t} + (\tfrac{1}{2}-A) e^{i\beta t} \right],
\end{equation}
where $\beta = \sqrt{(\pi \Delta_{\text{TLS}})^2 - (\gamma/2)^2}$ and the constant $A$ follows from the initial condition $\dot C(0) = -\gamma/4 + i(1-4A)\beta/2$, which depends on the initial probabilities for finding the qubit in the high- or low-frequency state.
Assuming equal probabilities at $t=0$, this condition becomes $\dot C_{\rm eq}(0) = 0$ and
\begin{equation}
    C_{\rm eq}(t) = \frac{1}{2} e^{-\gamma t / 2} \left[ \cos(\beta t) + \frac{\gamma}{2\beta} \sin(\beta t) \right].
\end{equation}
If the qubit is initially in the high- or low-frequency state, then the initial condition becomes $\dot C_\pm(0) = \pm \frac{i}{2}\pi\Delta_{\rm TLS}$, from which one finds $A_\pm = \frac{1}{4}[ 1 \mp(\pi\Delta_{\rm TLS}/\beta) + i(\gamma/2\beta) ]$, and the corresponding $C_\pm(t)$ follow from inserting this into Eq.~(\ref{eq:solcdo}).

The signal contrast available for state discrimination $S_{\text{AK}}(t) = |\Delta C(t)|$ is the absolute magnitude of the difference in coherence between the two TLS configurations $\Delta C(t) = C_+(t) - C_-(t)$, which follows from (\ref{eq:solcdo}) as
\begin{equation}
    \Delta C(t) = 2i(A_--A_+)e^{-\gamma t / 2}\sin(\beta t),
\end{equation}
and thus
\begin{equation}\label{eq:sak}
    S_{\text{AK}}(t) = \frac{\pi\Delta_{\text{TLS}}}{\beta} e^{-\gamma t / 2} |\sin(\beta t)|.
\end{equation}
We note that to derive (\ref{eq:sak}) one can also use the fact that $\Delta C(t) = -i2\dot{C}_{\rm eq}(t)/(\pi\Delta_{\rm TLS})$.
This relation arises directly from the coupled master equations for the partial coherences $C_\pm(t)$, where summing their time derivatives exactly cancels the stochastic jump terms, yielding $\dot{C}_{\rm eq} = (\dot{C}_+ + \dot{C}_-)/2 = i\pi\Delta_{\rm TLS}(C_+ - C_-)/2 = i\pi\Delta_{\rm TLS} \Delta C/2$.

To maximize the contrast, the system is probed at the optimal discrimination time $\tau_{\text{opt}} = 1/(2\Delta_{\text{TLS}})$. However, the true operational estimation bandwidth $\text{BW}$ is bottlenecked by the hardware overhead time $t_{\text{wall}}$ (readout and resonator reset), such that $\text{BW} = 1/(\tau_{\text{opt}} + t_{\text{wall}})$. 
Assuming the slow-switching approximation ($\gamma \ll 2\pi\Delta_{\text{TLS}}$) during the Ramsey sequence, we approximate $\beta \approx \pi\Delta_{\text{TLS}}$, reducing the Anderson-Kubo contrast to $S_{\text{AK}}(\tau_{\text{opt}}) \approx e^{-\gamma \tau_{\text{opt}} / 2}$. Furthermore, if the TLS switches state during the idle hardware overhead period $t_{\text{wall}}$, the estimation becomes outdated and subsequent active feedback fails. Combining the SPAM visibility $\alpha$, intrinsic decoherence $T_2$, the Anderson-Kubo phase-corruption during $\tau_{\text{opt}}$, and the TLS instability during $t_{\text{wall}}$, the total probability of a state assignment error is
\begin{equation}
    p_{\text{err}} \approx \frac{1}{2} \left[1 - \alpha e^{-\tilde \Gamma \tau_{\rm opt} - \gamma t_{\rm wall}} \right],
\end{equation}
with $\tilde \Gamma = 1/T_2 + \gamma/2$ the total effective dephasing rate.

\begin{figure}
    \centering
\includegraphics{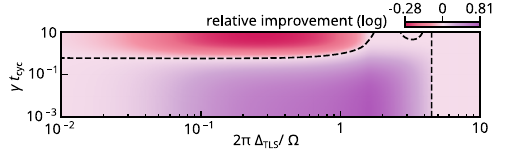}
\caption{Theoretical relative improvement of the X-gate infidelity using active estimation under the symmetric drive architecture ($\fc = (\fL+\fH)/2$), compared to the optimal average-frequency blind drive. The improvement is plotted as a function of the normalized TLS splitting $2\pi\Delta_{\rm TLS}/\Omega$ and the dimensionless switching rate $\gamma t_{\rm cyc}$. The dashed black contour marks the break-even boundary where no net improvement is obtained; in the upper red region, the active tracking loop is outpaced by the environmental switching, rendering it less effective than the static blind strategy.}
\label{fig:symmetric_plot}
\end{figure}

Substituting $\tau_{\text{opt}} = 1/(2\Delta_{\text{TLS}})$ and performing a first-order Taylor expansion reveals the linear additive error budget imposed by the hardware:
\begin{equation}
\label{eq:perr_BW}
    p_{\text{err}} \approx \frac{1}{2}\bigg[(1-\alpha) + \frac{\tilde \Gamma}{2\Delta_{\rm TLS}} + \gamma t_{\text{wall}}\bigg].
\end{equation}
This explicitly shows that the hardware overhead introduces a direct $\gamma t_{\text{wall}}$ penalty representing the probability of a fatal TLS flip occurring while the controller is tied up with measurement and reset tasks. This equation establishes the universal estimation error shared by both tracking architectures in the weak-noise limit. The relative improvement $\log_{10}[(1-F_{\rm blind})/(1-F_{\rm active})]$ mapped in Fig.~\ref{fig:3}(c) is computed by substituting this $p_{\text{err}}$ into Eq.~(\ref{eq:active_fidelity}), and comparing the resulting active fidelity against the respective blind baselines ($F_{\rm blind}^{\rm sym}$ and $F_{\rm blind}^{\rm asym}$) defined by the hardware constraints in Appendix~\ref{app:cons_est_error}. The parameters are set to $\alpha=2F_{\rm RO}-1=0.88$ (where $F_{\rm RO}=0.94$ is the readout assignment fidelity), $t_\pi = \SI{48}{\nano\second}$, $T_2 =\SI{61}{\micro\second}$ and $t_\text{\rm wall} = \SI{8}{\micro\second}$. From this simple theoretical model, we obtain an average improvement of $39\%$, in qualitative agreement with the measured $54\%$ average reduction in infidelity shown in Fig.~\ref{fig:3}(b).
For comparison, Fig.~\ref{fig:symmetric_plot} plots the symmetric architecture. Because its blind baseline optimizes around the average frequency, it establishes a lower error threshold. Consequently, when the estimation cycle becomes longer than the typical TLS switching time ($\gamma t_{\rm cyc} \gtrsim 1$), the active feedforward loop underperforms the static blind drive, resulting in a region of negative improvement.

\section*{References}
\bibliography{my_bibliography}

@article{gao2026non,
  title={Non-Local and Non-{Markovian} Effects of a Microscopic Two-Level Defect in Superconducting Quantum Circuits},
  author={Gao, Yang and Zhang, Yujia and Xu, Huikai and Shi, Pan and Li, Feiyu and Feng, Yaqing and Sun, Weijie and Ding, Jiayu and Liu, Yang and Wang, He and others},
  publisher={arXiv},    
  journal ="",
  year={2026},
  doi = {10.48550/arXiv.2605.23385}
}

@article{anderson1954,
  title = {A Mathematical Model for the Narrowing of Spectral Lines by Exchange or Motion},
  author = {Anderson, P. W.},
  journal = {Journal of the Physical Society of Japan},
  volume = {9},
  number = {3},
  pages = {316--339},
  year = {1954},
  doi = {10.1143/JPSJ.9.316}
}

@misc{Berritta2026Data,
  author = {Berritta, Fabrizio and
            Krzywda, Jan A. and
            Dvir, Tom and
            Buttles, Paul and
            Eilhart, Stanislav and
            Danon, Jeroen and
            Kuemmeth, Ferdinand},
  title = {Dataset underlying the manuscript:
           {Operating a bistable qubit}},
  howpublished = {Zenodo},
  year = {2026},
  doi = {10.5281/zenodo.23424770},
  url = {https://doi.org/10.5281/zenodo.23424770},
  note = {\url{https://doi.org/10.5281/zenodo.23424770}}
}

@article{kubo1954,
  title = {Note on the Stochastic Theory of Resonance Absorption},
  author = {Kubo, Ryogo},
  journal = {Journal of the Physical Society of Japan},
  volume = {9},
  number = {6},
  pages = {935--944},
  year = {1954},
  doi = {10.1143/JPSJ.9.935}
}

@article{chen2025,
      title={Scalable and Site-Specific Frequency Tuning of Two-Level System Defects in Superconducting Qubit Arrays}, 
      author={Larry Chen and Kan-Heng Lee and Chuan-Hong Liu and Brian Marinelli and Ravi K. Naik and Ziqi Kang and Noah Goss and Hyunseong Kim and David I. Santiago and Irfan Siddiqi},
publisher={arXiv},    
journal ="",
doi={10.48550/arXiv.2503.04702}, 
year={2025},
}

@article{roy2026two,
  title={Two-Level System Spectroscopy from Correlated Multilevel Relaxation in Superconducting Qubits},
  author={Roy, Tanay and You, Xinyuan and van Zanten, David and Crisa, Francesco and Garattoni, Sabrina and Zhu, Shaojiang and Grassellino, Anna and Romanenko, Alexander},
    publisher={arXiv},    
    journal ="",
doi={10.48550/arXiv.2602.11127},
year={2026}
}

@article{degnan2026reducing,
  title={Reducing {TLS} loss in tantalum {CPW} resonators using titanium sacrificial layers},
  author={Degnan, Zachary and Chiu, Chun-Ching and Chen, Yi-Hsun and Sommers, David and Abdurakhimov, Leonid and Zhu, Lihuang and Fedorov, Arkady and Jacobson, Peter},
    publisher={arXiv},    
    journal ="",
doi={10.48550/arXiv.2601.16369},
year={2026}
}

@article{marciniak2026,
      title={Millisecond-Scale Calibration and Benchmarking of Superconducting Qubits}, 
      author={Malthe A. Marciniak and Rune T. Birke and Johann B. Severin and Fabrizio Berritta and Daniel Kjær and Filip Nilsson and Smitha N. Themadath and Sangeeth Kallatt and James L. Webb and Kristoffer Bentsen and Tonny Madsen and Zhenhai Sun and Svend Krøjer and Christopher W. Warren and Jacob Hastrup and Morten Kjaergaard},
      publisher={arXiv},    
      journal ="",
      year={2026},
      doi={10.48550/arXiv.2602.11912}, 
}

@article{Berritta2025_T1,
  title = {Real-Time Adaptive Tracking of Fluctuating Relaxation Rates in Superconducting Qubits},
  author = {Berritta, Fabrizio and Benestad, Jacob and Krzywda, Jan A. and Krause, Oswin and Marciniak, Malthe A. and Kr\o{}jer, Svend and Warren, Christopher W. and Hogedal, Emil and Nylander, Andreas and Ahmad, Irshad and Osman, Amr and Bizn\'arov\'a, Janka and Rommel, Marcus and Roudsari, Anita Fadavi and Bylander, Jonas and Tancredi, Giovanna and Danon, Jeroen and Hastrup, Jacob and Kuemmeth, Ferdinand and Kjaergaard, Morten},
  journal = {Physical Review X},
  volume = {16},
  issue = {1},
  pages = {011025},
  numpages = {14},
  year = {2026},
  month = {Feb},
  publisher = {American Physical Society},
  doi = {10.1103/gk1b-stl3},
}

@article{luthi2018evolution,
  title={Evolution of nanowire transmon qubits and their coherence in a magnetic field},
  author={Luthi, Florian and Stavenga, Thijs and Enzing, OW and Bruno, Alessandro and Dickel, Christian and Langford, NK and Rol, Michiel Adriaan and Jespersen, Thomas Sand and Nyg{\aa}rd, Jesper and Krogstrup, P and others},
  journal={Physical review letters},
  volume={120},
  number={10},
  pages={100502},
  year={2018},
  publisher={APS}
}

@article{kam2024detrimental,
  title={Detrimental {non-Markovian} errors for surface code memory},
  author={Kam, John F and Gicev, Spiro and Modi, Kavan and Southwell, Angus and Usman, Muhammad},
publisher = {arXiv},  
journal="",
  year={2024},
doi = {10.48550/arXiv.2410.23779}
}

@article{pataki2024coherent,
  title={Coherent errors in stabilizer codes caused by quasistatic phase damping},
  author={Pataki, D{\'a}vid and M{\'a}rton, {\'A}ron and Asb{\'o}th, J{\'a}nos K and P{\'a}lyi, Andr{\'a}s},
  journal={Physical Review A},
  volume={110},
  number={1},
  pages={012417},
  year={2024},
  publisher={APS},
doi = {https://doi.org/10.1103/PhysRevA.110.012417}
}

@article{hakoshima2021relationship,
  title={Relationship between costs for quantum error mitigation and {non-Markovian} measures},
  author={Hakoshima, Hideaki and Matsuzaki, Yuichiro and Endo, Suguru},
  journal={Physical Review A},
  volume={103},
  number={1},
  pages={012611},
  year={2021},
  publisher={APS},
doi = {10.1103/PhysRevA.103.012611}
}

@article{gustavsson2012dynamical,
  title={Dynamical decoupling and dephasing in interacting two-level systems},
  author={Gustavsson, Simon and Yan, Fei and Bylander, Jonas and Yoshihara, Fumiki and Nakamura, Yasunobu and Orlando, Terry P and Oliver, William D},
  journal={Physical Review Letters},
  volume={109},
  number={1},
  pages={010502},
  year={2012},
  publisher={APS}
}

@article{wolff2026structural,
  title={Structural control of two-level defect density revealed by high-throughput correlative measurements of Josephson junctions},
  author={Wolff, Oliver F and Mantry, Harshvardhan and Raja, Rahim and Peng, Wei-Hsiang and Singirikonda, Kaushik and Lee, Seungkyun and Sudhaman, Shishir and Goncalves, Rafael and Huang, Pinshane Y and Kou, Angela and others},
    publisher={arXiv},    
journal ="",
      doi={10.48550/arXiv.2602.11469}, 
  year={2026}
}

@article{rojas2023spatial,
  title={Spatial noise correlations beyond nearest neighbors in 28 {Si/Si-Ge} spin qubits},
  author={Rojas-Arias, Juan S and Noiri, Akito and Stano, Peter and Nakajima, Takashi and Yoneda, Jun and Takeda, Kenta and Kobayashi, Takashi and Sammak, Amir and Scappucci, Giordano and Loss, Daniel and others},
  journal={Physical Review Applied},
  volume={20},
  number={5},
  pages={054024},
  year={2023},
  publisher={APS}
}

@article{Rojas2026,
  title = {Inferring Charge-Noise Source Locations from Correlations in Spin Qubits},
  author = {Rojas-Arias, J. S. and Noiri, A. and Yoneda, J. and Stano, P. and Nakajima, T. and Takeda, K. and Kobayashi, T. and Scappucci, G. and Tarucha, S. and Loss, D.},
  journal = {Physical Review Letters},
  volume = {136},
  issue = {2},
  pages = {027001},
  numpages = {7},
  year = {2026},
  month = {Jan},
  publisher = {American Physical Society},
  doi = {10.1103/glp5-bb5b},
}

@article{Zanuz2024,
  title = {Mitigating losses of superconducting qubits strongly coupled to defect modes},
  author = {Colao Zanuz, Dante and Ficheux, Quentin and Michaud, Laurent and Orekhov, Alexei and Hanke, Kilian and Flasby, Alexander and Bahrami Panah, Mohsen and Norris, Graham J. and Kerschbaum, Michael and Remm, Ants and Swiadek, Fran\cois and Hellings, Christoph and Laz\ifmmode \u{a}\else \u{a}\fi{}r, Stefania and Scarato, Colin and Lacroix, Nathan and others},
  journal = {Physical Review Applied},
  volume = {23},
  issue = {4},
  pages = {044054},
  numpages = {13},
  year = {2025},
  publisher = {American Physical Society},
  doi = {10.1103/PhysRevApplied.23.044054},
}

@article{weeden2025,
      title={Statistics of Strongly Coupled Defects in Superconducting Qubits}, 
      author={S. Weeden and D. C. Harrison and S. Patel and M. Snyder and E. J. Blackwell and G. Spahn and S. Abdullah and Y. Takeda and B. L. T. Plourde and J. M. Martinis and R. McDermott},
      year={2025},
    journal = "",
    publisher = {arXiv},
    doi={10.48550/arXiv.2506.00193} 
}

@article{kim2025error,
  title={Error mitigation with stabilized noise in superconducting quantum processors},
  author={Kim, Youngseok and Govia, Luke CG and Dane, Andrew and van den Berg, Ewout and Zajac, David M and Mitchell, Bradley and Liu, Yinyu and Balakrishnan, Karthik and Keefe, George and Stabile, Adam and others},
  journal={Nature Communications},
  volume={16},
  number={1},
  pages={8439},
  year={2025},
  publisher={Nature Publishing Group UK, London}
}

@article{dane2025,
      title={Performance Stabilization of High-Coherence Superconducting Qubits}, 
      author={Andrew Dane and Karthik Balakrishnan and Brent Wacaser and Li-Wen Hung and H. J. Mamin and Daniel Rugar and Robert M. Shelby and Conal Murray and Kenneth Rodbell and Jeffrey Sleight},
      year={2025},
      journal = "",
    publisher = {arXiv},
    doi= {10.48550/arXiv.2503.12514}
}

@article{dumoulin2024silicon,
    author = {Dumoulin Stuyck, Nard and Seedhouse, Amanda E. and Serrano, Santiago and Tanttu, Tuomo and Gilbert, Will and Huang, Jonathan Yue and Hudson, Fay and Itoh, Kohei M. and Laucht, Arne and Lim, Wee Han and Yang, Chih Hwan and Saraiva, Andre and Dzurak, Andrew S.},
    title = {Silicon spin qubit noise characterization using real-time feedback protocols and wavelet analysis},
    journal = {Applied Physics Letters},
    volume = {124},
    number = {11},
    pages = {114003},
    year = {2024},
    month = {03},
    doi = {10.1063/5.0179958},}

@article{mcrae2021,
author={McRae, Corey Rae H and Stiehl, Gregory M and Wang, Haozhi and Lin, Sheng-Xiang and Caldwell, Shane A and Pappas, David P and Mutus, Josh and Combes, Joshua},  
title={Reproducible coherence characterization of superconducting quantum devices},
  journal={Applied Physics Letters},
  volume={119},
  number={10},
pages= {100501},
  year={2021},
  publisher={AIP Publishing},
doi = {10.1063/5.0060370}
}

@article{Blais2021,
  title={Circuit quantum electrodynamics},
  author={Blais, Alexandre and Grimsmo, Arne L and Girvin, Steven M and Wallraff, Andreas},
  journal={Reviews of Modern Physics},
  volume={93},
  number={2},
  pages={025005},
  year={2021},
  publisher={APS},
doi = {https://doi.org/10.1103/RevModPhys.93.025005}
}

@article{Gebhart2023,
	title={Learning quantum systems},
	author={Gebhart, Valentin and Santagati, Raffaele and Gentile, Antonio Andrea and Gauger, Erik M and Craig, David and Ares, Natalia and Banchi, Leonardo and Marquardt, Florian and Pezz{\`e}, Luca and Bonato, Cristian},
	journal={Nature Reviews Physics},
	volume={5},
	number={3},
	pages={141--156},
	year={2023},
        doi = {https://doi.org/10.1038/s42254-022-00552-1},
	publisher={Nature Publishing Group UK London}
}

@Article{Arshad2024,
	author    = {Arshad, Muhammad Junaid and Bekker, Christiaan and Haylock, Ben and Skrzypczak, Krzysztof and White, Daniel and Griffiths, Benjamin and Gore, Joe and Morley, Gavin W. and Salter, Patrick and Smith, Jason and Zohar, Inbar and Finkler, Amit and Altmann, Yoann and Gauger, Erik M. and Bonato, Cristian},
	journal   = {Physical Review Applied},
	title     = {Real-time adaptive estimation of decoherence timescales for a single qubit},
	year      = {2024},
	number    = {2},
	pages     = {024026},
	volume    = {21},
	doi       = {10.1103/physrevapplied.21.024026},
	publisher = {American Physical Society (APS)},
}

@Article{Krantz2019,
    author = {Krantz, P. and Kjaergaard, M. and Yan, F. and Orlando, T. P. and Gustavsson, S. and Oliver, W. D.},
    title = {A quantum engineer's guide to superconducting qubits},
    journal = {Applied Physics Reviews},
    volume = {6},
    number = {2},
    pages = {021318},
    year = {2019},
    month = {06},
    doi = {10.1063/1.5089550},
}

@Article{Malley2015,
	author    = {O’Malley, P. J. J. and Kelly, J. and Barends, R. and Campbell, B. and Chen, Y. and Chen, Z. and Chiaro, B. and Dunsworth, A. and Fowler, A. G. and Hoi, I.-C. and Jeffrey, E. and Megrant, A. and Mutus, J. and Neill, C. and Quintana, C. and others},
	journal   = {Physical Review Applied},
	title     = {Qubit Metrology of Ultralow Phase Noise Using Randomized Benchmarking},
	year      = {2015},
	number    = {4},
	pages     = {044009},
	volume    = {3},
	doi       = {10.1103/physrevapplied.3.044009},
	publisher = {American Physical Society (APS)},
}

@article{Berritta2024b,
	title = {Physics-informed tracking of qubit fluctuations},
	author = {Berritta, Fabrizio and Krzywda, Jan A. and Benestad, Jacob and van der Heijden, Joost and Fedele, Federico and Fallahi, Saeed and Gardner, Geoffrey C. and Manfra, Michael J. and van Nieuwenburg, Evert and Danon, Jeroen and Chatterjee, Anasua and Kuemmeth, Ferdinand},
	journal = {Physical Review Applied},
	volume = {22},
	pages = {014033},
	numpages = {10},
	year = {2024},
	publisher = {American Physical Society},
	doi = {10.1103/PhysRevApplied.22.014033},
}

@Article{Berritta2024a,
	author    = {Berritta, Fabrizio and Rasmussen, Torbjørn and Krzywda, Jan A. and van der Heijden, Joost and Fedele, Federico and Fallahi, Saeed and Gardner, Geoffrey C. and Manfra, Michael J. and van Nieuwenburg, Evert and Danon, Jeroen and Chatterjee, Anasua and Kuemmeth, Ferdinand},
	journal   = {Nature Communications},
	title     = {Real-time two-axis control of a spin qubit},
	year      = {2024},
	pages    = {1676},
	volume    = {15},
	doi       = {10.1038/s41467-024-45857-0},
	publisher = {Springer Science and Business Media LLC},
}

@article{Berritta2025_FBS,
  title = {Efficient Qubit Calibration by Binary-Search {Hamiltonian} Tracking},
  author = {Berritta, Fabrizio and Benestad, Jacob and Pahl, Lukas and Mathews, Melvin and Krzywda, Jan A. and Assouly, R\'eouven and Sung, Youngkyu and Kim, David K. and Niedzielski, Bethany M. and Serniak, Kyle and Schwartz, Mollie E. and Yoder, Jonilyn L. and Chatterjee, Anasua and Grover, Jeffrey A. and Danon, Jeroen and Oliver, William D. and Kuemmeth, Ferdinand},
  journal = {PRX Quantum},
  volume = {6},
  issue = {3},
  pages = {030335},
  numpages = {11},
  year = {2025},
  publisher = {American Physical Society},
  doi = {10.1103/77qg-p68k},
}

@article{Shalibo2010,
  title = {Lifetime and Coherence of Two-Level Defects in a {Josephson} Junction},
  author = {Shalibo, Yoni and Rofe, Ya'ara and Shwa, David and Zeides, Felix and Neeley, Matthew and Martinis, John M. and Katz, Nadav},
  journal = {Physical Review Letters},
  volume = {105},
  issue = {17},
  pages = {177001},
  numpages = {4},
  year = {2010},
  publisher = {American Physical Society},
  doi = {10.1103/PhysRevLett.105.177001},
}

@Article{Schloer2019,
  author    = {Schlör, Steffen and Lisenfeld, Jürgen and Müller, Clemens and Bilmes, Alexander and Schneider, Andre and Pappas, David P. and Ustinov, Alexey V. and Weides, Martin},
  journal   = {Physical Review Letters},
  title     = {Correlating Decoherence in Transmon Qubits: Low Frequency Noise by Single Fluctuators},
  year      = {2019},
  issn      = {1079-7114},
  month     = nov,
  number    = {19},
  pages     = {190502},
  volume    = {123},
  doi       = {10.1103/physrevlett.123.190502},
  publisher = {American Physical Society (APS)},
}

@article{lisenfeld2019,
  title={Electric field spectroscopy of material defects in transmon qubits},
  author={Lisenfeld, J{\"u}rgen and Bilmes, Alexander and Megrant, Anthony and Barends, Rami and Kelly, Julian and Klimov, Paul and Weiss, Georg and Martinis, John M and Ustinov, Alexey V},
  journal={npj Quantum Information},
  volume={5},
  number={1},
  pages={105},
  year={2019},
  publisher={Nature Publishing Group UK London},
doi = {10.1038/s41534-019-0224-1}
}

@article{bilmes2020,
  title={Resolving the positions of defects in superconducting quantum bits},
  author={Bilmes, Alexander and Megrant, Anthony and Klimov, Paul and Weiss, Georg and Martinis, John M and Ustinov, Alexey V and Lisenfeld, J{\"u}rgen},
  journal={Scientific reports},
  volume={10},
  number={1},
  pages={3090},
  year={2020},
  publisher={Nature Publishing Group UK London},
doi = {10.1038/s41598-020-59749-y}
}

@article{grabovskij2012,
  title={Strain tuning of individual atomic tunneling systems detected by a superconducting qubit},
  author={Grabovskij, Grigorij J and Peichl, Torben and Lisenfeld, J{\"u}rgen and Weiss, Georg and Ustinov, Alexey V},
  journal={Science},
  volume={338},
  number={6104},
  pages={232--234},
  year={2012},
  publisher={American Association for the Advancement of Science},
doi = {10.1126/science.1226487}
}

@Article{Park2025,
  author    = {Park, Jaemin and Jang, Hyeongyu and Sohn, Hanseo and Yun, Jonginn and Song, Younguk and Kang, Byungwoo and Stehouwer, Lucas E. A. and Esposti, Davide Degli and Scappucci, Giordano and Kim, Dohun},
  journal   = {Nature Communications},
  title     = {Passive and active suppression of transduced noise in silicon spin qubits},
  year      = {2025},
  month     = jan,
  pages    = {78},
  volume    = {16},
  doi       = {10.1038/s41467-024-55338-z},
  publisher = {Springer Science and Business Media LLC},
}

@article{gertler2021protecting,
  title={Protecting a bosonic qubit with autonomous quantum error correction},
  author={Gertler, Jeffrey M and Baker, Brian and Li, Juliang and Shirol, Shruti and Koch, Jens and Wang, Chen},
  journal={Nature},
  volume={590},
  number={7845},
  pages={243--248},
  year={2021},
  publisher={Nature Publishing Group UK London}
}

@article{levine2024demonstrating,
  title={Demonstrating a long-coherence dual-rail erasure qubit using tunable transmons},
  author={Levine, Harry and Haim, Arbel and Hung, Jimmy SC and Alidoust, Nasser and Kalaee, Mahmoud and DeLorenzo, Laura and Wollack, E Alex and Arrangoiz-Arriola, Patricio and Khalajhedayati, Amirhossein and Sanil, Rohan and others},
  journal={Physical Review X},
  volume={14},
  number={1},
  pages={011051},
  year={2024},
  publisher={APS}
}

@article{szankowski2017environmental,
  title={Environmental noise spectroscopy with qubits subjected to dynamical decoupling},
  author={Sza{\'n}kowski, Piotr and Ramon, Guy and Krzywda, Jan and Kwiatkowski, Damian and others},
  journal={Journal of Physics: Condensed Matter},
  volume={29},
  number={33},
  pages={333001},
  year={2017},
  publisher={IOP Publishing}
}

@article{ye2025stabilizing,
  title={Stabilizing an individual charge fluctuator in a {Si/Si-Ge} quantum dot},
  author={Ye, Feiyang and Ellaboudy, Ammar and Nichol, John M},
  journal={Physical Review Applied},
  volume={23},
  number={4},
  pages={044063},
  year={2025},
  publisher={APS}
}

@article{ye2024,
  title={Characterization of individual charge fluctuators in {Si/SiGe} quantum dots},
  author={Ye, Feiyang and Ellaboudy, Ammar and Albrecht, Dylan and Vudatha, Rohith and Jacobson, N Tobias and Nichol, John M},
  journal={Physical Review B},
  volume={110},
  number={23},
  pages={235305},
  year={2024},
  publisher={APS},
doi = {10.1103/PhysRevB.110.235305}
}

@article{donnelly2025noise,
  title={Noise correlations in an atom-based quantum dot array},
  author={Donnelly, MB and Rowlands, J and Kranz, L and Hsueh, YL and Chung, Y and Timofeev, AV and Geng, H and Singh-Gregory, P and Gorman, SK and Keizer, JG and others},
  journal={Physical Review Applied},
  volume={23},
  number={6},
  pages={064058},
  year={2025},
  publisher={APS}
}

@article{liu2024observation,
  title={Observation of discrete charge states of a coherent two-level system in a superconducting qubit},
  author={Liu, Bao-Jie and Wang, Ying-Ying and Sheffer, Tal and Wang, Chen},
  journal={Physical Review Letters},
  volume={133},
  number={16},
  pages={160602},
  year={2024},
  publisher={APS},
  doi ={https://doi.org/10.1103/PhysRevLett.133.160602}
}

@article{hutin2025preparing,
  title={Preparing {Schr{\"o}dinger} cat states in a microwave cavity using a neural network},
  author={Hutin, Hector and Bilous, Pavlo and Ye, Chengzhi and Abdollahi, Sepideh and Cros, Loris and Dvir, Tom and Shah, Tirth and Cohen, Yonatan and Bienfait, Audrey and Marquardt, Florian and others},
  journal={PRX Quantum},
  volume={6},
  number={1},
  pages={010321},
  year={2025},
  publisher={APS}
}

@Article{McKay2017,
	author    = {McKay, David C. and Wood, Christopher J. and Sheldon, Sarah and Chow, Jerry M. and Gambetta, Jay M.},
	journal   = {Physical Review A},
	title     = {Efficient {Z} gates for quantum computing},
	year      = {2017},
	number    = {2},
	pages     = {022330},
	volume    = {96},
	doi       = {10.1103/physreva.96.022330},
	publisher = {American Physical Society (APS)},
}

@Article{Knill2008,
	author    = {Knill, E. and Leibfried, D. and Reichle, R. and Britton, J. and Blakestad, R. B. and Jost, J. D. and Langer, C. and Ozeri, R. and Seidelin, S. and Wineland, D. J.},
	journal   = {Physical Review A},
	title     = {Randomized benchmarking of quantum gates},
	year      = {2008},
	number    = {1},
	pages     = {012307},
	volume    = {77},
	doi       = {10.1103/physreva.77.012307},
	publisher = {American Physical Society (APS)},
}

@article{muller2019,
  title={Towards understanding two-level-systems in amorphous solids: insights from quantum circuits},
  author={M{\"u}ller, Clemens and Cole, Jared H and Lisenfeld, J{\"u}rgen},
  journal={Reports on Progress in Physics},
  volume={82},
  number={12},
  pages={124501},
  year={2019},
  publisher={IOP Publishing},
doi = {10.1088/1361-6633/ab3a7e}
}

@article{Rist2013,
  title         = {
    Millisecond charge-parity fluctuations and induced decoherence in a superconducting transmon qubit
  },
  author        = {
    Ristè, D. and Bultink, C. C. and Tiggelman, M. J. and Schouten, R. N. and
    Lehnert, K. W. and DiCarlo, L.
  },
  year          = {2013},
  month         = may,
  journal       = {Nature Communications},
  publisher     = {Springer Science and Business Media LLC},
  volume        = {4},
  pages        = {1913},
  doi           = {10.1038/ncomms2936},
  issn          = {2041-1723},
}

@Article{Siddiqi2021,
  author    = {Siddiqi, Irfan},
  journal   = {Nature Reviews Materials},
  title     = {Engineering high-coherence superconducting qubits},
  year      = {2021},
  issn      = {2058-8437},
  number    = {10},
  pages     = {875--891},
  volume    = {6},
  doi       = {10.1038/s41578-021-00370-4},
  publisher = {Springer Science and Business Media LLC},
}

@Article{Murray2021,
  author    = {Murray, Conal E.},
  journal   = {Materials Science and Engineering: R: Reports},
  title     = {Material matters in superconducting qubits},
  year      = {2021},
  issn      = {0927-796X},
  month     = oct,
  pages     = {100646},
  volume    = {146},
  doi       = {10.1016/j.mser.2021.100646},
  publisher = {Elsevier BV},
}
\end{document}